# Multiscale simulations of growth-dominated $Sb_2Te$ phase-change material for non-volatile photonic applications

Xu-Dong Wang[1#], Wen Zhou[1#], Hangming Zhang[1], Shehzad Ahmed[1], Tiankuo Huang[1], Riccardo Mazzarello[2], En Ma[1], Wei Zhang[1*]

[1] Center for Alloy Innovation and Design (CAID), State Key Laboratory for Mechanical Behavior of Materials, Xi'an Jiaotong University, Xi'an, 710049, China.
[2] Department of Physics, Sapienza University of Rome, Rome, 00185, Italy

[#]These authors contributed equally to this work.

*Email: wzhang0@mail.xjtu.edu.cn

**Abstract**

Chalcogenide phase-change materials (PCMs) are widely applied in electronic and photonic applications, such as non-volatile memory and neuro-inspired computing. Doped $Sb_2Te$ alloys are now gaining increasing attention for on-chip photonic applications, due to their growth-driven crystallization features. However, it remains unknown whether $Sb_2Te$ also forms a metastable crystalline phase upon nanoseconds crystallization in devices, similar to the case of nucleation-driven Ge-Sb-Te alloys. Here, we carry out ab initio simulations to understand the changes in optical properties of amorphous $Sb_2Te$ upon crystallization and post annealing. During the continuous transformation process, changes in the dielectric function are highly wavelength-dependent from the visible-light range towards the telecommunication band. Our finite-difference time-domain simulations based on the ab initio input reveal key differences in device output for color display and photonic memory applications upon tellurium ordering. Our work serves as an example of how multiscale simulations of materials can guide practical photonic phase-change applications.

**Introduction**

Chalcogenide phase change materials (PCMs) can rapidly and reversibly switch between crystalline and amorphous phases. The large difference in atomic structure and bonding mechanism between the two phases gives rise to a notable contrast in electrical and optical properties, which can be used to encode digital data[1-10]. The invention of the very first successful product based on PCMs can be traced back to the 1990s, when rewritable optical media (CD, DVD and Blu-ray Disc) entered the market as a game changer[11-13]. In the recent decade, an emerging type of non-volatile electronic memory utilizing PCMs has been developed and commercialized with superior device performance in switching speed, storage capacity and scalability[14-19]. Thanks to the fast development of silicon waveguide platforms, PCM-based optical and photonic devices are gaining renewed attention, since they can go beyond the diffraction limit for high-density integration. Several advanced applications have been demonstrated, including non-volatile photonic memory[20-24], neuromorphic computing[25-30], flexible displays and color rendering[31-35], metasurfaces and low-loss applications[36-42].

Two families of chalcogenide alloys have been used in commercial products, i.e., the pseudo-binary GeTe-$Sb_2Te_3$ alloys, in particular, $Ge_2Sb_2Te_5$ (GST)[11], and Ag and In doped $Sb_2Te$ (AIST)[12]. The major difference between these two families of alloys is the nucleation rate. The nucleation time for GST (of the order of nanoseconds)[43-45] is more than three orders of magnitude faster than that of AIST (of order microseconds)[46-48]. The difference in the crystallization mode (nucleation or growth) in binary Sb-Te system as a function of Sb:Te ratio was explained in Ref. [49]. In typical memory devices of sub-micrometer scale, the crystallization of GST occurs via incubation of multiple nuclei and subsequent grain growth (nucleation-type), while AIST typically crystallizes via the forefront of amorphous/crystalline interfaces (growth-type). Upon device miniaturization, crystallization of growth-type PCMs can also be completed in nanoseconds level, because of the high interface growth rate[48,50]; an example being doped $Sb_2Te$[51,52]. Importantly, their crystallization process can be better controlled leading to superior programming consistencies, because of the absence of randomness associated with the stochastic nucleation process[53]. Therefore, growth-type PCMs are attracting increasing attention for photonic applications recently[54-56].

Recently, it was shown that the crystallization temperature $T_c$ of amorphous PCMs can be significantly increased by scaling the thickness down to a few nanometers, exploiting the nanoconfinement effects[57-60]. Even amorphous Sb thin films (3 nm) can be stabilized at 20 °C over 50 hours[57], in spite of the fact that the $T_c$ of amorphous Sb is well below room temperature. In this way, doping is no longer compulsory to enhance the amorphous stability for practical applications. Here, we focus on the prototypical growth-dominated alloy $Sb_2Te$, which shows a fair $T_c$ of ~150 °C[61] in ~100 nm thick thin films. One can expect that in ultra-thin films of $Sb_2Te$ (e.g. ~5 nm) the stability of the amorphous state will be further enhanced so as to make them suitable for most non-volatile photonic applications Importantly, the contrast in optical properties is mainly determined by the parent alloy, $Sb_2Te$, not by impurities[56]. However, in-depth atomic-scale understanding of $Sb_2Te$ upon rapid crystallization at elevated temperatures is still lacking. In this work, we carry out multiscale simulations on $Sb_2Te$, including density functional (DFT) based ab initio molecular dynamics (AIMD) calculations and finite-difference time-domain (FDTD) simulations, to unveil the structural and optical evolution of $Sb_2Te$ during ultrafast crystallization and subsequent thermal annealing. We reveal that amorphous $Sb_2Te$ could crystallize into a disordered rhombohedral structure, showing rather different electronic and optical properties with respect to the A7 ground state. Our multiscale point towards different strategies for device programming due to the formation of metastable crystalline $Sb_2Te$ at the nanoscale.

**Results**
**Rapid crystallization and subsequent atomic ordering**
The ground state of $Sb_2Te$ is in rhombohedral A7 structure as shown in Figure 1a. Its primitive cell is composed of one $Sb_2Te_3$ quintuple-layer and two $Sb_2$ bilayer blocks, which are alternately stacked and separated by van der Waals-like gaps. To obtain the crystallized structure via rapid crystal growth, we adopted a similar AIMD simulation protocol with fixed crystalline layers, as done in our previous work[50,62]. We built a large model of crystalline $Sb_2Te$ in an orthorhombic supercell with the size of 2.24 × 2.16 × 5.16 nm$^3$ (540 Sb atoms and 270 Te atoms). Each atomic layer has 30 atoms. One Sb layer and one Te layer were kept fixed during the melt-quench calculation. Specifically, the model was first randomized at 3000 K, and then quenched down to 1000 K in 6 ps. It was held at 1000 K for ~30 ps and then quenched to 300 K with a cooling rate of 25 K/ps to create a planar amorphous/crystalline interface. Subsequently, the model was reheated and annealed at ~600 K, and became fully crystallized in ~135 ps. See more technical details in the Methods section. Figure 1b shows some snapshots of the crystallization trajectory. Obviously, the crystallized structure is different from the fully ordered A7 structure, in that the Te atoms are not arranged in layers but scattered in the bulk, which can be attributed to the fast atomic diffusion and large sticking coefficients[50].

We took several atomic configurations during the rapid growth process and quenched them down very rapidly down to zero K by direct geometry relaxation. These partially crystallized states are non-volatile, and could in principle be obtained in devices by sending weak or short pulses. The corresponding total energy of these intermediate states calculated at zero K is shown in Figure 2a, indicating a continuous energy reduction by ~150 meV/atom. Yet, this crystallized state is still about ~65 meV/atom higher in energy as compared to the ground state A7 structure. Post annealing could in principle reduce this energy difference and drive the transformation to the more ordered ground state. This trend is consistent with the case of GST, where a metastable rocksalt-like phase containing high amount of atomic disorder is firstly formed upon nanoseconds crystallization, and is then gradually transformed into an ordered trigonal phase (with a similar energy reduction of ~68 meV/atom) upon further long-term thermal annealing via vacancy ordering[63-68]. Since the timescale of the post annealing process (minutes to hours) is beyond the reach of ab initio simulations, we manually swapped nearest-neighboring Sb and Te atoms to mimic the post annealing process, which can be regarded as a Te ordering process, i.e., the scattered Te in the bulk are rearranged into layers until the perfect ordered A7 structure is obtained. To quantify this structural ordering process, we define a parameter $I_{Te}$ as the atomic concentration of Te atoms in the Te-rich layers. The larger the $I_{Te}$, the higher structural similarity with respect to A7 structure. More specifically, the model with fully random distribution of Te atoms corresponds to $I_{Te}$=33.3%, while that with fully ordered Te layers corresponds to $I_{Te}$=100% (A7). The $I_{Te}$ value of the crystallized structure is approximately 46.7%. Selected structures with different level of Te ordering can be found in Figure 2c, in which the green stripes highlight the Te-rich layers. Clearly, the energy is gradually reduced upon Te ordering.

In comparison with our previous work on AIST[50], the crystallization simulation of the same size then

(it took ~170 ps) is slower than Sb$_2$Te, due to the presence of dopants. The simulated crystallized phase of AIST was also in a disordered rhombohedral structure, consistent with experimental observation upon rapid laser irradiation[69]. Atomic ordering into layered structures with clear separation of Sb$_2$Te$_3$ blocks with In and Ag impurities and Sb$_2$ slabs was also observed in AIST thin films upon long-term thermal annealing at elevated temperatures[70]. Hence, our work resolves the discrepancy in experimental observations by separating the rapid crystallization process obtained on the nanosecond timescale with the subsequent atomic ordering in the crystalline phase upon long-term thermal annealing on the minute timescale. Next, we evaluated the impact of crystallization and Te ordering on dielectric function, which is important for optical applications using Sb$_2$Te-based alloys.

**Evolution of optical properties**

Next, we carried out DFT calculations to evaluate the change in optical properties during the crystallization and Te ordering process of Sb$_2$Te. All atomic structures were fully relaxed at zero K prior to these calculations. The obtained real ($\varepsilon_1$) and imaginary ($\varepsilon_2$) parts of the dielectric function are shown in Figure 3a and 3b. The spectrum is chosen to range from 400 to 2000 nm, covering the visible light for the non-volatile display application and the telecom wavelength for photonic memory/computing applications. As crystallization proceeds, $\varepsilon_1$ decreases in the whole spectrum, while $\varepsilon_2$ increases in the long wavelength region (1200–2000 nm) but slightly decreases in the visible light region (400–800 nm). For the Te ordering process, both $\varepsilon_1$ and $\varepsilon_2$ show small variations in the visible light region but sizable changes in the long wavelength region. To see the trends more clearly, we averaged the $\varepsilon_1$ and $\varepsilon_2$ values in the range 400–800 nm and 1400–1800 nm for both crystallization and Te ordering process, as shown in Figure 3c. The variation in the visible light range turns out to be smaller than that in the long wavelength region. It is interesting to note that the Te ordering process greatly reduces the optical contrast window between amorphous and crystallized states in the long wavelength region for both $\varepsilon_1$ and $\varepsilon_2$. This behavior is different from that of GST alloys, whose vacancy ordering process after crystallization further enlarges the contrast window[65,71]. However, the Te ordering process has a much smaller effect on the visible light region.

To further validate the optical variations for the Te ordering process, we built five independent models for each $I_{Te}$ value by using a random number generator, and calculated their optical properties. The models were built by randomly swapping Sb/Te atomic pairs in the A7 structure (Figure 4a), rather than only swapping the nearest-neighbor pairs (Figure 2c). As shown in Figure 4b and Supplementary Figure 1, the dielectric function profiles of these five models are highly consistent with that shown in Figure 3. The changes in dielectric functions have direct influences on the refractive index ($n$), extinction coefficient ($k$) and reflectivity coefficient ($R$) that are exploited for applications. These quantities were using the following equations[72]:

$$(1) \quad n(\omega) = \left(\frac{\sqrt{\varepsilon_1^2 + \varepsilon_2^2} + \varepsilon_1}{2}\right)^{\frac{1}{2}},$$

$$(2) \quad k(\omega) = \left(\frac{\sqrt{\varepsilon_1^2 + \varepsilon_2^2} - \varepsilon_1}{2}\right)^{\frac{1}{2}},$$

$$(3) \quad R(\omega) = \frac{(n-1)^2 + k^2}{(n+1)^2 + k^2}.$$

As shown in Figure 4c-e, these three optical profiles show smaller variations in the visible light region but much larger changes in the telecom band upon Te ordering process. This implies that the annealing process after crystallization should have a large impact on the memory/computing applications, while probably a small influence on the non-volatile display applications.

To explain this trend, we plotted the total density of states (TDOS) profiles for each $I_{Te}$ value in Figure 5a. As the $I_{Te}$ value decreases, the TDOS values from −1.0 to 0.5 eV (with respect to the Fermi energy, which is shifted as zero) increase, while the values within 0.5–1.0 eV become smaller. The TDOS profiles beyond these two regions show marginal variations. These findings are also confirmed by hybrid functional calculations, which generally give more accurate predictions of the electronic states near the Fermi energy (Supplementary Figure 2). The same trend also holds for the optical calculations (Supplementary Figure 3). The TDOS curves and the corresponding inverse participation ratio (IPR) values of all disordered models are included in Supplementary Figure 4 and Supplementary Figure 5. Although the A7 phase is not a semiconductor, it shows a pseudo gap near the Fermi level. As the $I_{Te}$ value decreases, this pseudo gap is gradually filled, resulting in more free carriers for light absorption in the disordered crystalline models (Figure 4d). The change in TDOS is also mirrored in the joint density of states (JDOS), which is one of the key components for $\varepsilon_2$, accounting for the amount of possible inter-band excitations. The other key component for $\varepsilon_2$ is the transition dipole moment (TDM), quantifying the transition probability for each possible excitation[73-75]. As shown in Figure 5b, the JDOS values increase with the reduction of $I_{Te}$ in the long wavelength (low photon energy) region, which leads to the increase in $\varepsilon_2$ values. However, for the small wavelength (high photon energy) region, the JDOS curves are almost unaffected by variations in $I_{Te}$. Therefore, the decrease of $\varepsilon_2$ values in this region is expected to originate from the variations of TDM, which indeed decreases with the reduction of $I_{Te}$ in the visible light regime (Figure 5c).

Furthermore, we investigated the structural origin for the variation of the TDOS profiles. We plotted the partial DOS (PDOS) per atom for both Te and Sb atoms in Figure 6a, showing similar trend as the TDOS in Figure 5a. This indicates that both types of atoms contribute to the variations of the electronic structure. The PDOS curves of all disordered models are included in Supplementary Figure 4. We selected the $I_{Te}$=86.7% model as an example, and further analyzed the local DOS (LDOS) of all the atoms in terms of their coordination environment, as shown in Figure 6b. We defined $CN_{Te}$ as the number of Te atoms surrounding the center atom, see Figure 6c. The $CN_{Te}$ value is zero for all the Te atoms, and is either 0 (within the $Sb_2$ slabs), 3 (at the interface) or 6 (within the $Sb_2Te_3$ block) for Sb atoms in the ordered A7 structure. However, for the disordered structures, such as the structure of the $I_{Te}$=86.7%, the $CN_{Te}$ value can range from 0 to 6 for both Te and Sb atoms (Supplementary Figure 6). In this model, a clear increase in LDOS at and above the Fermi level is observed as $CN_{Te}$ increases for both Te and Sb atoms. The trend is stronger for Te atoms because more valence electrons can be found locally due to the enriched Te-Te bonds[62]. This behavior is also reflected in the calculated IPR values (Supplementary Figure 5). Regarding the A7 model with no Te-Te bonds, the LDOS for Te atoms is nearly vanished at the Fermi level, similar to the case of $CN_{Te}$ = 0 of the disordered model. While for Sb atoms, there are three neighboring configurations, which show a nearly vanishing LDOS at the Fermi level (comparable to $CN_{Te}$=0 of the disordered model) but a sizeable LDOS around 0.5 eV (appearing between $CN_{Te}$=3 and 6 of the disordered model).

**Non-volatile display devices**

We then propose two potential photonic applications of $Sb_2Te$ using FDTD simulations (see Methods)[76]. So far, optical experiments of $Sb_2Te$ are still limited, and it also remains unclear whether a metastable configuration can be obtained upon rapid crystallization using femtosecond or picosecond laser pulses. Hence, we used the calculated $n$, $k$ and $R$ from ab initio simulations as the input parameters for the FDTD simulations. First, we consider non-volatile display applications. Figure 7a shows the schematic of a reflective display thin-film device, incorporating indium tin oxide (ITO)/$Sb_2Te$/ITO multilayer on top of a silver mirror coated on a glass substrate. The thickness of each layer has been chosen to enhance the color contrast between the amorphous and crystalline phases. Specifically, $h_{ITO1}$, $h_{Sb2Te}$, $h_{ITO2}$ are 10, 15, 130 nm, respectively. These thin films can generate resonant reflections based on the optical cavity effect[31], which are determined by the phase of the $Sb_2Te$ thin film in the cavity. Figure 7b shows the simulated reflectance spectra of the designed thin-film device with the amorphous (amor.), crystallized (cryst.) and fully ordered (A7) $Sb_2Te$. Obviously, the reflectance curves of crystallized and fully ordered structures almost overlap, differing notably from that of the amorphous phase. As shown in Figure 7c, we mapped the reflectance spectra onto the CIE1931 chromaticity diagram by using the CIE color-matching equation[77], and showed that the colors of the two crystalline structures are indeed close to each other. We further converted the spectra to the RGB color code[77], and simulated the university logo encoded on the multilayer devices as shown in Figures 7d and 7e, exhibiting excellent display performance. Such display application can be realized by either electrical pulse stimulation or laser direct write[31-34], and a recent study has demonstrated experimentally that $Sb_2Te$ thin films can be used for color rendering[55]. Our simulations suggest that further thermal annealing of the crystallized phase would not alter the encoded color significantly, which guarantees the robustness of the display performance.

**Waveguide devices for photonic memory and computing**
Another potential application for $Sb_2Te$ is the photonic memory and computing. Figure 8a shows the schematic of a silicon-on-insulator (SOI) strip waveguide decorated with $Sb_2Te$ and ITO thin films. ITO is a protection layer used to prevent oxidation of $Sb_2Te$. Thickness and width of the SOI waveguide are denoted as $h_{wg}$ and $w_{wg}$, respectively, and those of the $Sb_2Te$ (ITO) thin film are denoted as $h_{Sb2Te}$ and $d_{Sb2Te}$ ($h_{ITO}$ and $d_{ITO}$). Transmittance ($T$) of light passing through the waveguide device is $P_2/P_1$, where $P_1$ and $P_2$ are the power of incident and transmitted light, respectively. We investigated transmittance of light with two sets of geometrical parameters: $h_{wg}$=0.34 μm, $w_{wg}$=0.5 μm, $h_{Sb2Te}$=5 nm, $h_{ITO}$=10 nm, and $d_{Sb2Te}$=$d_{ITO}$=2 or 1 μm. The incident light is with the fundamental transverse magnetic (TM) mode and a wavelength range of 1500–1600 nm. Figure 8b shows the transmission spectra of the $Sb_2Te$ waveguide device with varied phases. Specifically, at a wavelength of 1560 nm, waveguide transmittances are respectively 25.03%, 0.96%, and 11.31% for the amorphous, crystallized, and A7 phases for the device with $d_{Sb2Te}$=2 μm. These values can be further enlarged by reducing $d_{Sb2Te}$ to 1 μm, giving 50.30%, 9.31%, and 33.90%, respectively.

The contrast window of the transmittance between the amorphous and crystallized phases can reach 40% for $Sb_2Te$, which is larger than that of the devices based on GST alloys (~20%)[21], showing its potential for photonic memory and computing. As compared with the amorphous and fully ordered phases, the largely reduced output light intensity of the crystallized state is due to its much higher light absorption (larger $k$), which is also mirrored in the simulated electric field |$E$| profiles in Figure 8c-e. Importantly, we note that upon further annealing after crystallization, the contrast window of

transmittance is not enlarged but reduced as highlighted by the arrows in Figure 8b, which is different from the case of GST alloys[71]. Moreover, the geometrical parameters of the waveguide device can be further optimized to enhance the on-off ratio of the light transmission, which is desirable for practical applications. It is also noted that the sizable contrast in optical properties between the metastable rocksalt phase and the stable hexagonal phase of GST or $Sb_2Te_3$ alloys has been utilized for non-volatile optical applications[78-80], which could in principle bypass the aging issues of the amorphous phase[81]. It remains to be explored whether the relatively large change in transmittance between the crystallized and A7 structure of $Sb_2Te$ with $d_{Sb2Te}$ = 1 μm can also be exploited for optical applications.

**Discussion**

We found that the atomic structure of crystalline $Sb_2Te$ obtained after ultrafast crystallization is a disordered A7 structure with Te atoms randomly distributed in the bulk. The crystallized structure can be further annealed to induce a tellurium ordering process, and gradually transformed to a fully ordered structure with all the Te atoms forming perfect layers in the $Sb_2Te_3$ blocks. The structural difference between the rapidly crystallized and post annealed structures leads to a difference in electronic structure and thereby wavelength-dependent variations of the optical properties, i.e., it has notable influence in the telecom region, but negligible influence in the visible light region. Our multiscale simulations suggest that once programmed, the $Sb_2Te$-based display devices are insensitive to potential temperature fluctuations in the crystalline phase. In the ultrascaled devices of a few nm thickness, structural relaxation of the amorphous phase is also prohibited[7]. Therefore, $Sb_2Te$-based alloys could potentially lead to superior device performance for photonic applications that utilize the visible-light range. Regarding the memory/computing applications that rely on the telecom band, we suggest that additional temperature rise should be avoided after programming, and high-speed and high-frequency device operations could be desirable to gain the large programming window for $Sb_2Te$-based devices. Our work showcases how atomic understanding of materials can guide device design, and is expected to stimulate future exploration of $Sb_2Te$-based alloys for various photonic phase-change applications.

**Methods**

*Ab initio calculations*. We carried out DFT-based ab initio molecular dynamics (AIMD) simulations by using the second-generation Car–Parrinello molecular dynamics scheme[82] as implemented in the CP2K package[83]. A triple-zeta plus polarization Gaussian-type basis set was used to expand the Kohn–Sham orbitals, and plane waves with a cutoff of 300 Ry were used to expand the charge density. Scalar-relativistic Goedecker-Teter-Hutter (GTH) pseudopotentials[84] and Perdew–Burke–Ernzerhof (PBE) functional[85] were applied in the calculations. The simulations were conducted in the canonical ensemble (NVT), and the temperature was controlled by a stochastic Langevin thermostat with the time step of 2 fs. The Vienna Ab-initio Simulation Package (VASP)[86] was used to compute the dielectric functions after structural relaxation of the models. PBE functional and PAW pseudopotentials[87] were used with the energy cutoff value set as 400 eV. Γ point was used to sample the Brillouin zone of the models with 810 atoms. The frequency-dependent dielectric matrix was calculated within the independent-particle approximation. Neglecting local field effects and many body effects was proven to be adequate to quantify the optical contrast between crystalline and amorphous PCMs[73-75].

*FDTD simulations*. The reflective display thin-film device was modeled by 2D finite-difference time-domain (FDTD) simulation (Lumerical FDTD Solutions)[76]. Build-in models of refractive indices were used for silver, glass, and silicon. The light source is a plane wave (placed at $z$ = 1 μm) with the propagation direction along –$z$ axis, which is perpendicular to the thin-film surface. A monitor was placed above the light source (at $z$ = 1.5 μm) to detect reflectance spectra. Perfectly matched layer (PML) boundary condition was applied at $z$ = ±2 μm. Periodic boundary condition was applied for the rest boundaries. The symmetric geometry of the thin-film device enables polarization-insensitive operation. The photonic waveguide memory devices were modeled by 3D FDTD with the fundamental transverse magnetic (TM) mode injected into a waveguide. A light source was placed at $x$ = −0.5 μm and a monitor was placed at $x$ = 4 μm to detect transmission spectra. PML boundary condition was applied to all boundaries. 3D FDTD simulation also calculated electric field profiles |$E$| at $y$ = 0 plane. The grid sizes were set fine enough to obtain converged simulation results. Especially, Δ$z$ was set as 1 nm for the $Sb_2Te$ and ITO thin films.

## DATA AVAILABILITY
The data that support the findings of this study are available from the corresponding author, Professor Wei Zhang (email: wzhang0@mail.xjtu.edu.cn), upon reasonable request.

## CODE AVAILABILITY
The authors declare that the applied software supporting the findings of this study are commercially available in the VASP software package https://www.vasp.at and Lumerical FDTD Solutions https://www.ansys.com/products/photonics/fdtd.

## ACKNOWLEDGEMENTS
E.M. acknowledges the National Natural Science Foundation of China (Grant no. 52150710545). W.Z. thanks the support of the International Joint Laboratory for Micro/Nano Manufacturing and Measurement Technologies of Xi'an Jiaotong University. W.Z. and E.M. are grateful to XJTU for the support of their work at CAID. The authors acknowledge the computational resources provided by the HPC platform of Xi'an Jiaotong University, the National Supercomputing Center in Xi'an, and the Hefei Advanced Computing Center.

## AUTHOR CONTRIBUTIONS
W. Zhang conceptualized the work. The ab initio simulations were performed by X.-D.W. with contributions from H.Z. and S.A. The FDTD simulations were performed by W. Zhou with contributions from T.H. The paper was written by X.-D.W., W. Zhou and W. Zhang with contributions from R.M. and E.M. All authors contributed to the discussions and analyses of the data, and approved the final version.

## COMPETING INTERESTS
The authors declare no competing interests.

## ADDITIONAL INFORMATION
Supplementary information The online version contains supplementary material available at

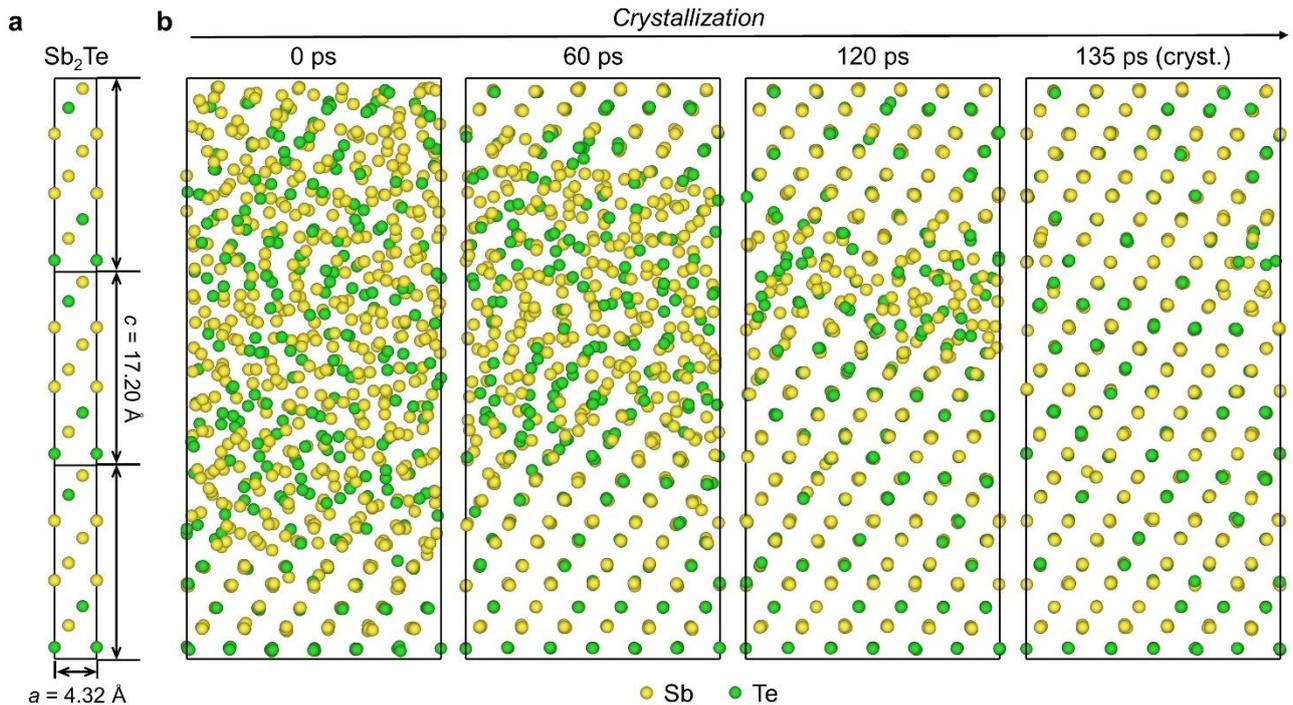

**Fig. 1. Structural evolution during simulated crystallization of Sb$_2$Te at ~600 K. a** The most stable phase of Sb$_2$Te in fully ordered A7 structure with alternating Sb$_2$Te$_3$ and Sb$_2$ atomic blocks. **b** Snapshots of the crystallization process initiated by the amorphous/crystalline interface. The model is fully crystallized after 135 ps (denoted as "cryst."), resulting in a disordered structure with Te atoms randomly distributed in the bulk.

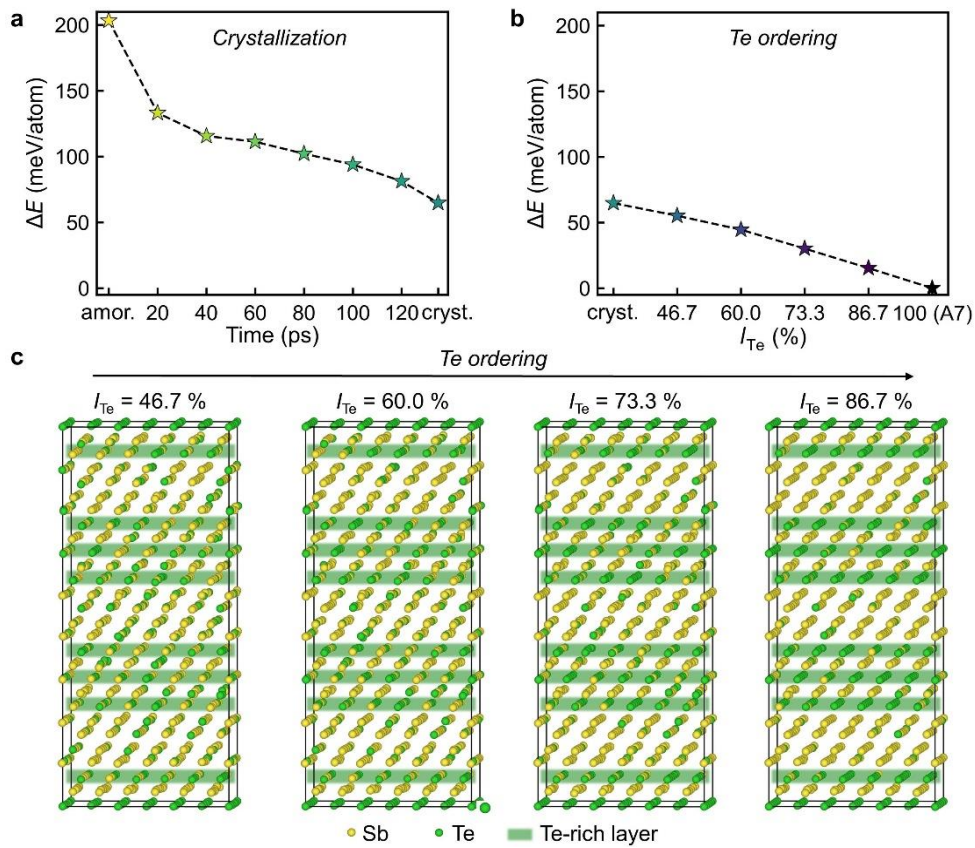

**Fig. 2. Energy profiles for the crystallization and subsequent Te ordering processes. a** Energy profile for the crystallization process of $Sb_2Te$. The amorphous, crystallized and intermediate structures were optimized at zero K for energy data collection. **b** Energy profile for the Te ordering process after crystallization. All the energy data points were obtained at zero K. **c** Snapshots of Te ordering process. The models are generated by manually swapping nearest-neighboring Sb and Te atoms. $I_{Te}$ indicates the concentration of Te atoms in the Te-rich layers highlighted by green stripes.

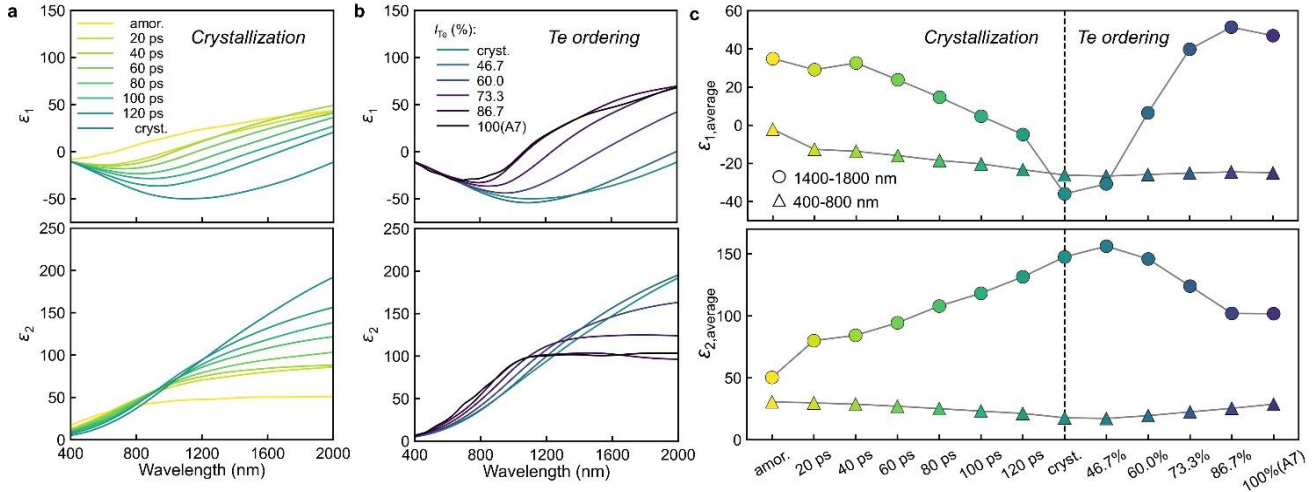

**Fig. 3. Dielectric function.** The real part ($\varepsilon_1$) and imaginary part ($\varepsilon_2$) of the dielectric function during **a** the crystallization process and **b** the Te ordering process. **c** The averaged $\varepsilon_1$ and $\varepsilon_2$ over the spectrum range of 1400-1800 nm and 400-800 nm.

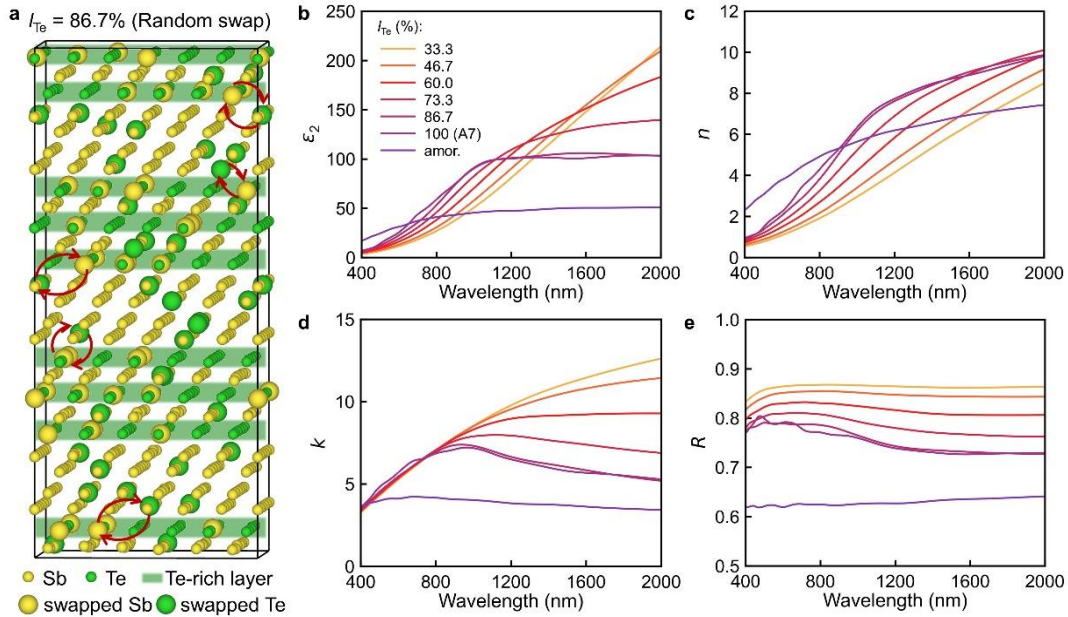

**Fig. 4. Optical variations for random-swap models. a** Example of a "random-swap" model with $I_{Te}$=86.7%. Red arrows highlight the random-swapping process between Sb/Te pairs. The swapped atoms that deviate from the fully ordered structure (Figure 1a) are depicted with larger spheres. **b-e** Imaginary part of the dielectric function ($\varepsilon_2$), refractive index ($n$), extinction coefficient ($k$) and reflectivity ($R$) of the random-swap models ranging from $I_{Te}$=33.3% to $I_{Te}$=100%. The data of the amorphous phase were added for a direct comparison.

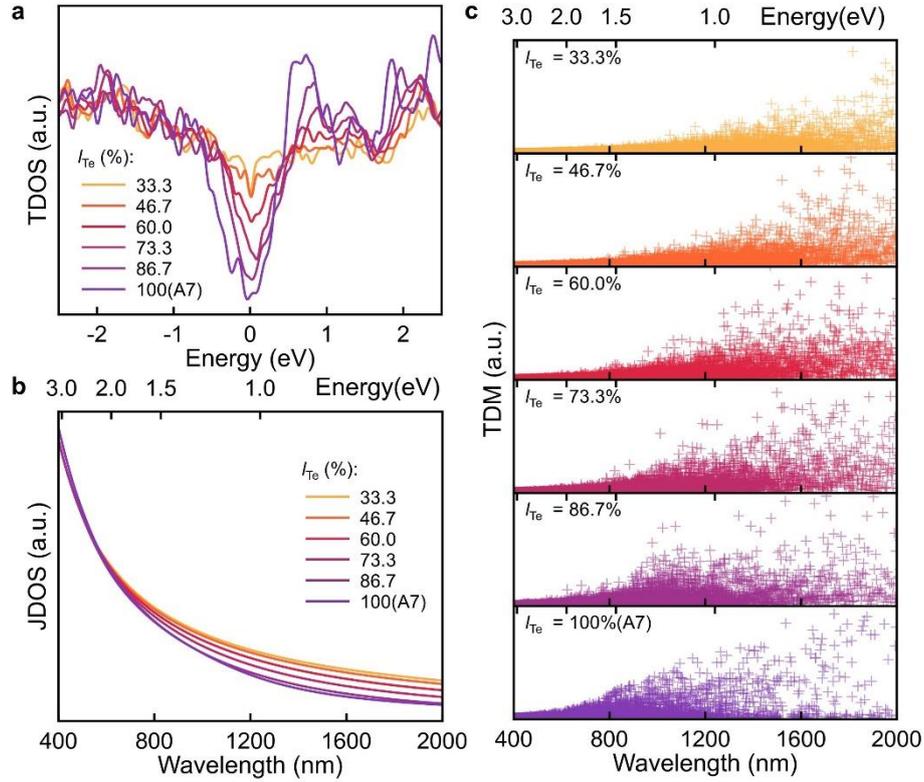

**Fig. 5. Electronic structure of random-swap models. a-c** Total density of states (TDOS), joint density of states (JDOS) and transition dipole moment (TDM) of the random-swap models ranging from $I_{Te}$=33.3% to $I_{Te}$=100%.

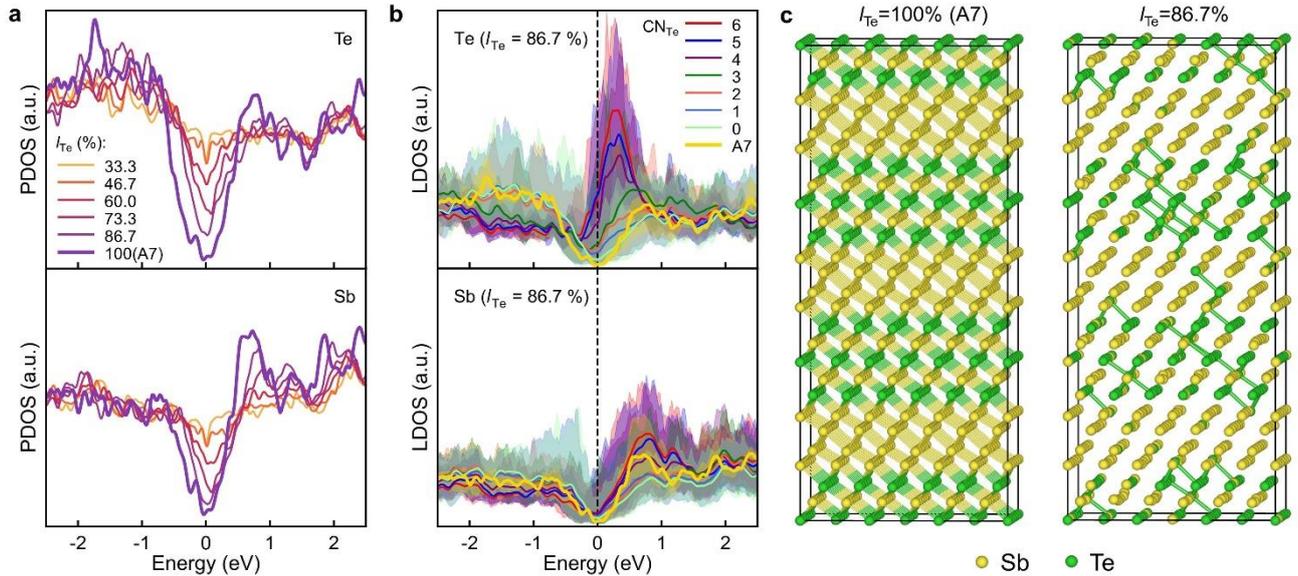

**Fig. 6. Relationship between atomic structure and electronic structure. a** Partial density of states (PDOS) per atom for Te and Sb atoms upon Te ordering. **b** Local density of states (LDOS) in terms of the coordination number $CN_{Te}$ for a model with $I_{Te}$=86.7% and the perfect A7 model. The shaded areas and solid curves are the envelope and averaged LDOS of all the atoms with the same $CN_{Te}$, respectively.

Note that the CN_Te value for Te atoms equals zero in the A7 model. **c** Models of A7 and $l_{Te}$=86.7% with atomic bonds being highlighted. For the latter model, only Te-Te homopolar bonds are shown.

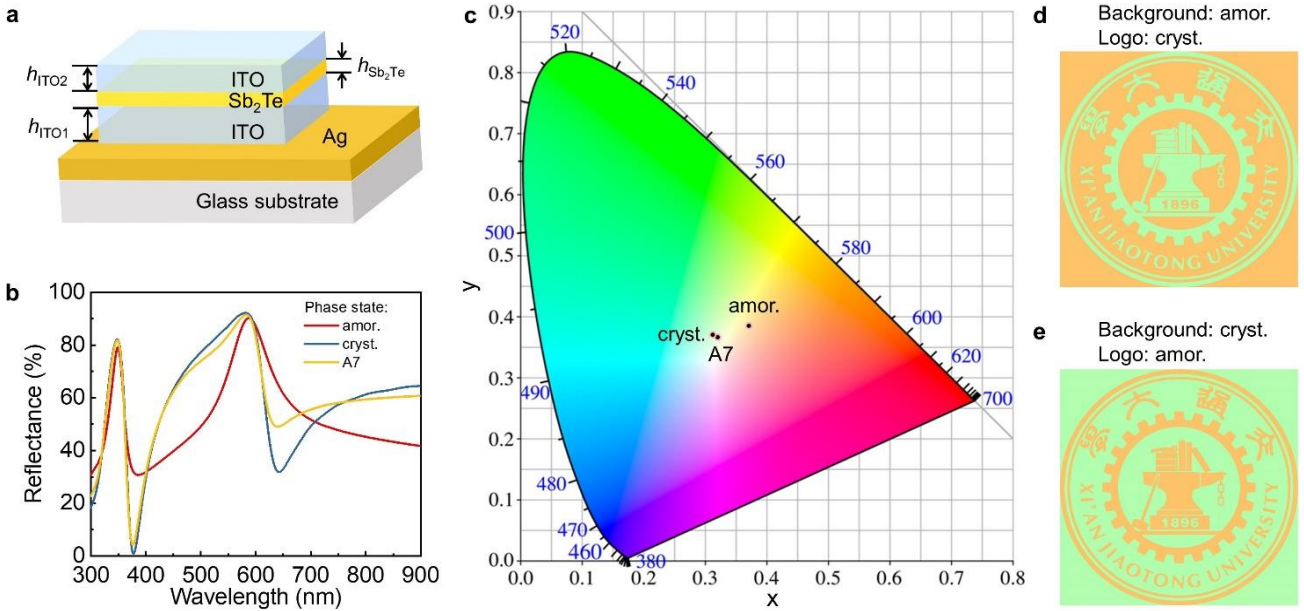

**Fig. 7. Multilayer thin film display. a** Schematic of a reflective display thin-film device, incorporating ITO/Sb$_2$Te/ITO multilayer on top of a silver mirror coated on a glass substrate. **b** Simulated reflectance spectra of the thin-film device in the amorphous (amor.), crystallized (cryst.) and fully ordered (A7) states. $h_{ITO1}$, $h_{SbTe}$ and $h_{ITO2}$ are 10, 15, 130 nm, respectively. **c** Color mapping of the three states on the CIE1931 chromaticity diagram. **d-e** Display of the university logo by exploiting the amorphous and crystallized states.

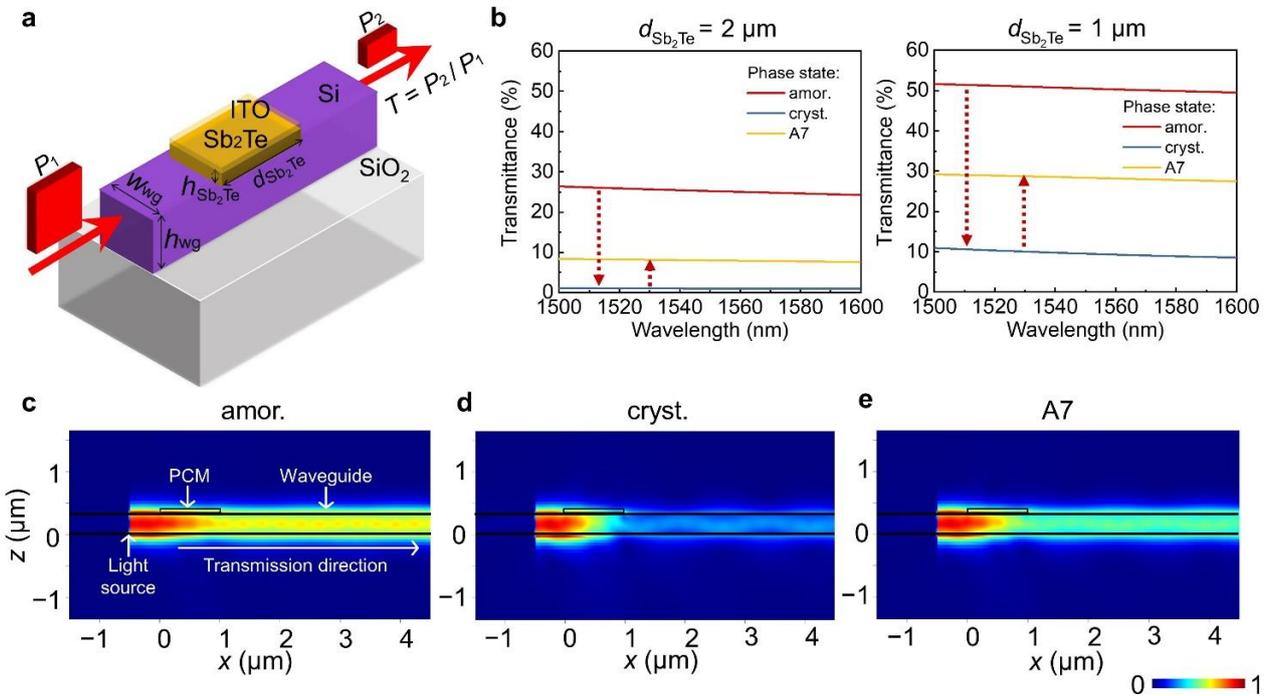

**Fig. 8. Photonic waveguide memory. a** Schematic of a SOI strip waveguide decorated with Sb$_2$Te and ITO thin films. **b** Transmission spectra of the waveguide device in the amorphous (amor.), crystallized (cryst.) and fully ordered (A7) states, with the length of Sb$_2$Te patch ($d_{Sb2Te}$) equal to 2 or 1 μm. **c-e** Simulated electric field $|E|$ profiles of the amor., cryst., and A7 states, respectively.

Supplementary Information for

# Multiscale simulations of growth-dominated Sb$_2$Te phase-change material for non-volatile photonic applications

Xu-Dong Wang[1#], Wen Zhou[1#], Hangming Zhang[1], Shehzad Ahmed[1], Tiankuo Huang[1],

Riccardo Mazzarello[2], En Ma[1], Wei Zhang[1]*

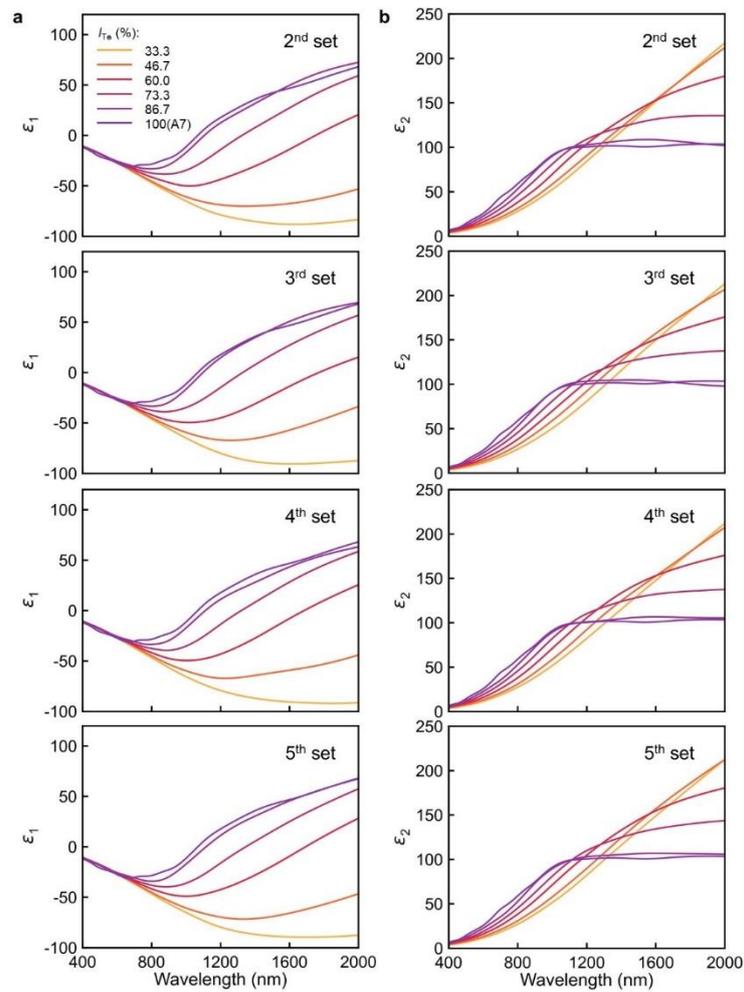

**Supplementary Figure 1. Dielectric functions. a** Real part ($\varepsilon_1$) and **b** imaginary part ($\varepsilon_2$) of dielectric functions of four other sets of models with random swapping.

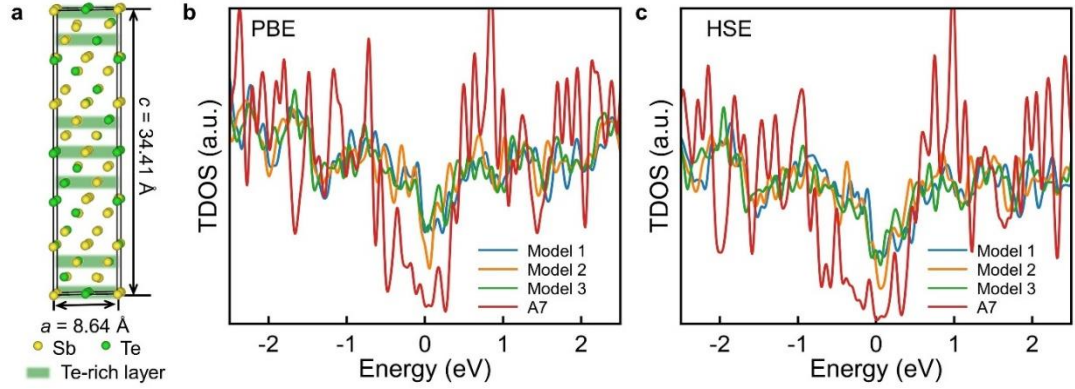

**Supplementary Figure 2. Hybrid functional calculations for total density of states (TDOS).** **a** The typical model used for hybrid functional (HSE06) calculation. This model was built in a hexagonal supercell with 72 atoms. The cell parameters are $a=b=8.64$ Å and $c=34.41$ Å. **b-c** TDOS profiles calculated by PBE and HSE06 functionals. Three independent models of $I_{Te}=50\%$ were built as a representative random swap model, denoting as Model 1-3. The TDOS curves of fully ordered structure (A7) are also included for comparison. The TDOS profiles of PBE and HSE functionals show consistent contrast between the random swap model and fully ordered model, in line of that shown in Figure 5a.

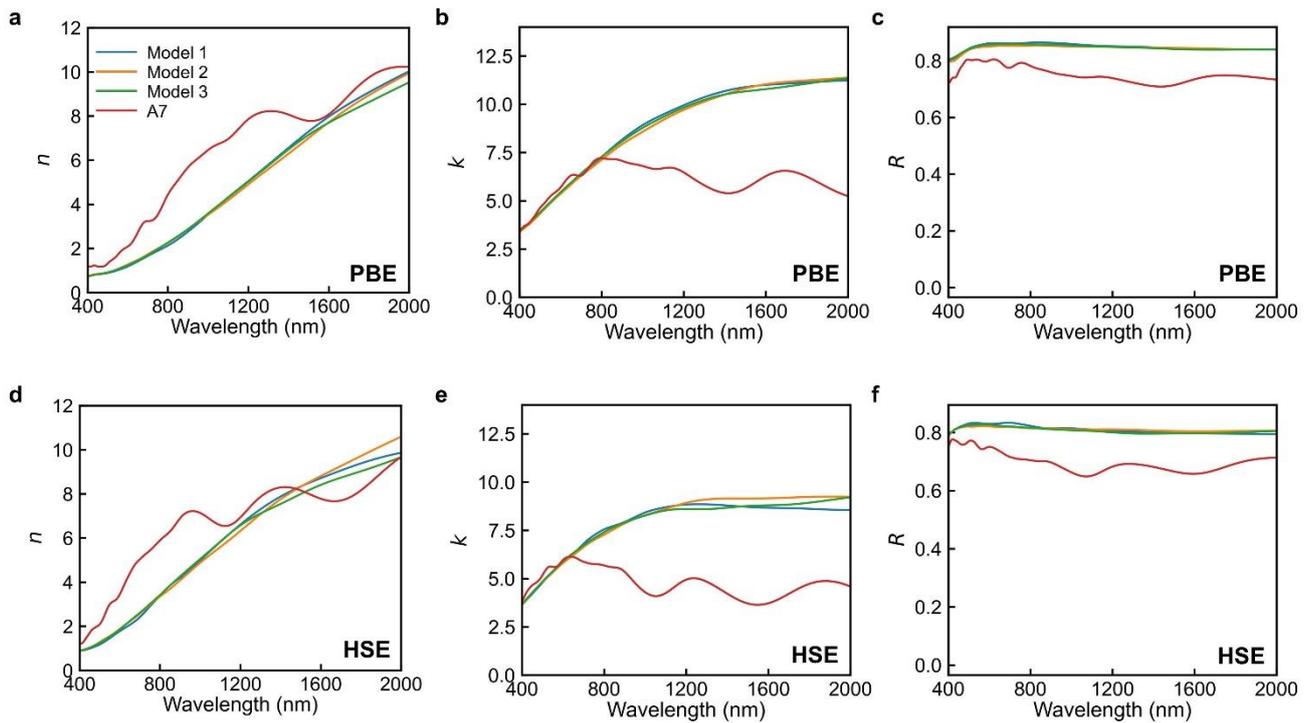

**Supplementary Figure 3. Hybrid functional calculations for optical properties.** **a-c** The calculated $n$, $k$ and $R$ for the A7 unit cell model and three disordered models ($I_{Te}=50\%$) shown in Supplementary Figure 2b with PBE functional. **d-f** The corresponding optical data calculated with hybrid functional.

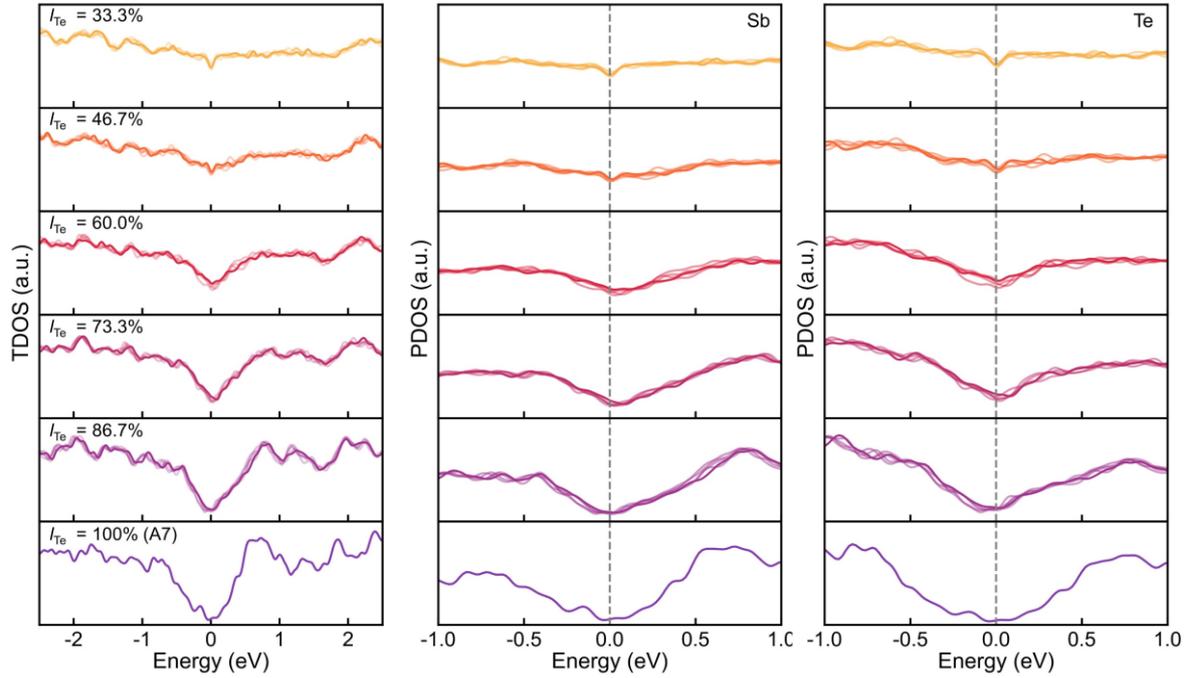

**Supplementary Figure 4. Statistical calculations.** The calculated TDOS and PDOS shown in Figure 5a and Figure 6a are plotted as solid lines, and the DOS curves of the statistical models are plotted in semi-transparent colors.

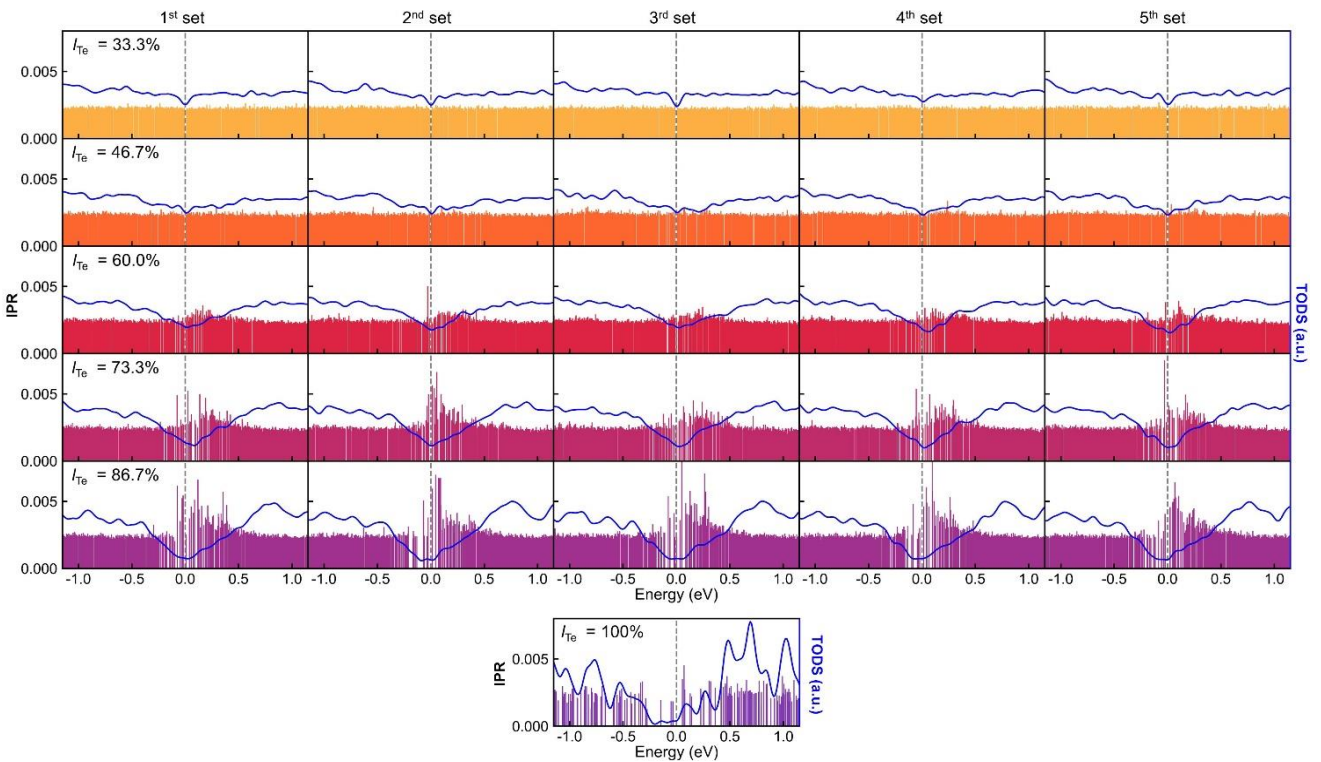

**Supplementary Figure 5. The inverse participation ration (IPR) analysis.** The IPR values of the A7 model were computed using a supercell of the same system size, as compared to the statistical models.

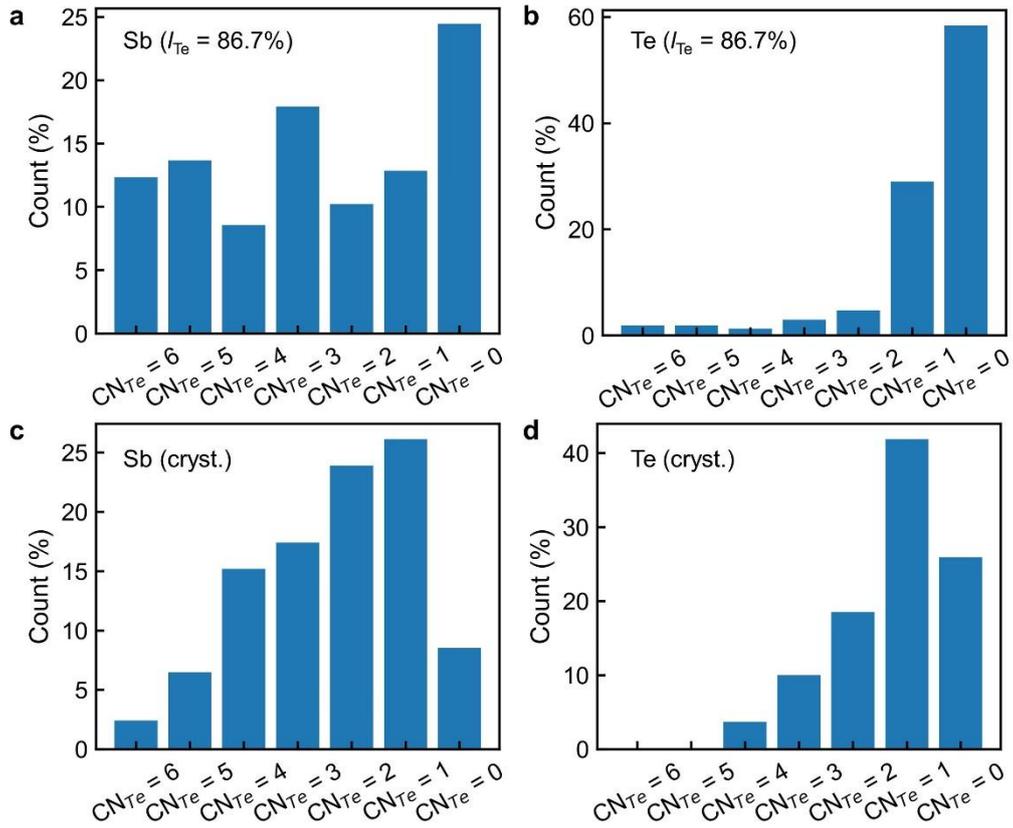

**Supplementary Figure 6. CN$_{Te}$ distributions. a-b** and **c-d** are the CN$_{Te}$ distributions for the $I_{Te}$=86.7% model and crystallized model, respectively.